\documentclass{article}
\usepackage{spconf}
\usepackage[noadjust]{cite}
\usepackage{bbm}
\usepackage{array}
\usepackage{dcolumn}
\usepackage{epsfig}
\usepackage[intlimits]{amsmath}
\usepackage{amsmath, amsfonts, yhmath, bm}
\usepackage{amssymb}
\usepackage{psfrag}
\usepackage{color,soul}
\usepackage[dvipsnames]{xcolor}
\usepackage[normalem]{ulem}
\usepackage{enumerate}
\usepackage{stackengine}
\usepackage{graphicx}
\usepackage{subcaption}
\usepackage{multirow}
\usepackage{etoolbox}
\usepackage{tcolorbox}
\usepackage{float}
\usepackage{dsfont}
\usepackage[ruled,vlined]{algorithm2e}
\usepackage{algpseudocode}
\usepackage{tikz}
\usepackage{mathtools}
\usepackage{arydshln}
\usetikzlibrary{shapes.geometric, arrows}

\usepackage{svg}
\usepackage{lipsum}  

\usepackage{hyperref}
\hypersetup{
	colorlinks,
	linkcolor={red!50!black},
	citecolor={blue!50!black},
	urlcolor={blue!80!black}
}

\usepackage{balance}
\usepackage{url}
\usepackage[inline]{enumitem} % For inline enumerations

\usepackage{vmr-symbols-vecbold}
\usepackage{standard-macros}
\PassOptionsToPackage{hyphens}{url}
\usepackage{hyperref}

\DeclareSymbolFontAlphabet{\amsmathbb}{AMSb}%

\newcommand{\lro}[1]{\lefto({#1}\right)}																% left right paranthesis operator
\newcommand{\lr}[1]{\left({#1}\right)}																% left right paranthesis operator
\newcommand{\lrb}[1]{\left \lbrace {#1} \right \rbrace}															% left right braces operator
\safemath{\dopplerspread}{B_D}																								% doppler spread
\safemath{\delayspread}{T_D}																									% delay spread
\safemath{\nc}{n\sub{c}}																										% coherence time
\safemath{\nf}{n\sub{f}}																										% feedback message length
\safemath{\efa}{p\sub{sc}}
\safemath{\efb}{p\sub{cs}}
\safemath{\ef}{\epsilon\sub{f}	}
\safemath{\nd}{n\sub{d}}																										% data symbols
\safemath{\ntx}{n\sub{t}} 																											% transmit antennas
\safemath{\nrx}{n\sub{r}}																											% receive antennas
\safemath{\ntxt}{\tilde{n\sub{t}}}																											% receive antennas
\safemath{\cb}{\ensuremath{L}} 																								% code blocks
\safemath{\cl}{\ensuremath{n}} 																								% codelength
\safemath{\txanto}{{\ensuremath{\tilde{m}_t}}} 																		% transmit antennas when some is turned off
\safemath{\cs}{M} 																														% code size
\safemath{\idPustm}{\ensuremath{S_{k}}}
\safemath{\error}{\ensuremath{\epsilon}} 																				%Error target
\safemath{\eexp}{\ensuremath{\mathcal{E}}} 																			%Error exponent
\safemath{\nsubc}{n\sub{s}}			 																						% number of subcarriers
\safemath{\nofdm}{n\sub{o}} 																									% number of OFDM symbols
\safemath{\bc}{\ensuremath{B_c}} 																							% Coherence bandwidth
\safemath{\ts}{\ensuremath{T_s}} 																							% Symbol time
\safemath{\nrb}{\ensuremath{n_{rb}}} 																						% Symbol time
\safemath{\rul}{\ensuremath{\rho\sub{ul}}}
\safemath{\rdl}{\ensuremath{\rho\sub{dl}}}

\safemath{\nres}{\ell}
\safemath{\nr}{n\sub{r}}
\safemath{\maxk}{M^*\lr{\nres, \nsubc, \nofdm, \epsilon, \rho}}
\safemath{\Rmax}{R^*}%\lr{\nres, \nsubc, \nofdm,M, \epsilon, \rho}}
\safemath{\Emin}{E\sub{b}^*/N_0}%\lr{\nres, \nsubc, \nofdm,M, \epsilon, \rho}}
\safemath{\Eminf}{\frac{E\sub{b}^*}{N_0}}
\safemath{\np}{\ensuremath{n\sub{p}}}
\safemath{\ndf}{\ensuremath{\bar{n}\sub{d}}}
\safemath{\npf}{\ensuremath{\bar{n}\sub{p}}}
\safemath{\code}{\ensuremath{\mathcal{C}}}
\safemath{\err}{\ensuremath{\epsilon}}
\safemath{\rp}{\ensuremath{\rho\sub{p}}}
\safemath{\rd}{\ensuremath{\rho\sub{d}}}
\safemath{\cohtime}{\ensuremath{T\sub{c}}}
\safemath{\cohbw}{\ensuremath{B\sub{c}}}
\safemath{\nmax}{\ensuremath{\ell\sub{m}}}
\safemath{\ntot}{\ensuremath{n\sub{tot}}}
\safemath{\nul}{\ensuremath{n\sub{ul}}}
\safemath{\ndl}{\ensuremath{n\sub{dl}}}

\safemath{\yp}{\ensuremath{\randvecy_{\nu}^{(\text{p})}}}
\safemath{\yd}{\ensuremath{\randvecy_{\nu}^{(\text{d})}}}
\safemath{\ypd}{\ensuremath{\vecy_{\nu}^{(\text{p})}}}
\safemath{\ydd}{\ensuremath{\vecy_{\nu}^{(\text{d})}}}

\safemath{\ypf}{\ensuremath{\bar{\randvecy}_{\nu}^{(\text{p})}}}
\safemath{\ydf}{\ensuremath{\bar{\randvecy}_{\nu}^{(\text{d})}}}
\safemath{\ypdf}{\ensuremath{\bar{\vecy}_{\nu}^{(\text{p})}}}
\safemath{\yddf}{\ensuremath{\bar{\vecy}_{\nu}^{(\text{d})}}}

\safemath{\xp}{\ensuremath{\vecx^{(\text{p})}}}
\safemath{\xd}{\ensuremath{\randvecx_{\nu}^{(\text{d})}}}
\safemath{\xdd}{\ensuremath{\vecx_{\nu}^{(\text{d})}}}

\safemath{\xpf}{\ensuremath{\bar{\vecx}^{(\text{p})}}}
\safemath{\xdf}{\ensuremath{\bar{\randvecx}_{\nu}^{(\text{d})}}}
\safemath{\xddf}{\ensuremath{\bar{\vecx}_{\nu}^{(\text{d})}}}

\safemath{\xdb}{\ensuremath{\overline{\randvecx}^{(\text{d})}}}
\safemath{\Pxd}{\ensuremath{P_{\randvecx^{(\text{d})}}}}

\safemath{\xpbar}{\ensuremath{\overline{\matX}^{(\text{p})}}}
\safemath{\xdbar}{\ensuremath{\overline{\randmatX}^{(\text{d})}}}

\safemath{\xdv}{\ensuremath{\randvecx^{(\text{d})}}}
\safemath{\xdbarv}{\ensuremath{\overline{\randvecx}^{(\text{d})}}}
\safemath{\ydv}{\ensuremath{\randvecy^{(\text{d})}}}

\safemath{\xdr}{\ensuremath{\matX^{(\text{d})}}}

\safemath{\ttx}{\ensuremath{\tau\sub{tx}}}
\safemath{\trx}{\ensuremath{\tau\sub{rx}}}
\safemath{\ack}{\ensuremath{\mathrm{s}}}
\safemath{\nack}{\ensuremath{\mathrm{c}}}

\newcommand{\prob}[1]{\ensuremath{\mathbb{P}\lro{#1}}}

\safemath{\mI}{\ensuremath{i\lro{\randvecy ; \randvecx}}} 				% i(Y;X)
\safemath{\randveca}{\bm{A}}
\safemath{\randvecb}{\bm{B}}
\safemath{\randvecc}{\bm{C}}
\safemath{\randvecd}{\bm{D}}
\safemath{\randvece}{\bm{E}}
\safemath{\randvecf}{\bm{F}}
\safemath{\randvecg}{\bm{G}}
\safemath{\randvech}{\bm{H}}
\safemath{\randveci}{\bm{I}}
\safemath{\randvecj}{\bm{J}}
\safemath{\randveck}{\bm{K}}
\safemath{\randvecl}{\bm{L}}
\safemath{\randvecm}{\bm{M}}
\safemath{\randvecn}{\bm{N}}
\safemath{\randveco}{\bm{O}}
\safemath{\randvecp}{\bm{P}}
\safemath{\randvecq}{\bm{Q}}
\safemath{\randvecr}{\bm{R}}
\safemath{\randvecs}{\bm{S}}
\safemath{\randvect}{\bm{T}}
\safemath{\randvecu}{\bm{U}}
\safemath{\randvecv}{\bm{V}}
\safemath{\randvecw}{\bm{W}}
\safemath{\randvecx}{\bm{X}}
\safemath{\randvecy}{\bm{Y}}
\safemath{\randvecz}{\bm{Z}}
\safemath{\randvecphi}{\bm{\Phi}}

\safemath{\randmatA}{\amsmathbb{A}}
\safemath{\randmatB}{\amsmathbb{B}}
\safemath{\randmatC}{\amsmathbb{C}}
\safemath{\randmatD}{\amsmathbb{D}}
\safemath{\randmatE}{\amsmathbb{E}}
\safemath{\randmatF}{\amsmathbb{F}}
\safemath{\randmatG}{\amsmathbb{G}}
\safemath{\randmatH}{\amsmathbb{H}}
\safemath{\randmatI}{\amsmathbb{I}}
\safemath{\randmatJ}{\amsmathbb{J}}
\safemath{\randmatK}{\amsmathbb{K}}
\safemath{\randmatL}{\amsmathbb{L}}
\safemath{\randmatM}{\amsmathbb{M}}
\safemath{\randmatN}{\amsmathbb{N}}
\safemath{\randmatO}{\amsmathbb{O}}
\safemath{\randmatP}{\amsmathbb{P}}
\safemath{\randmatQ}{\amsmathbb{Q}}
\safemath{\randmatR}{\amsmathbb{R}}
\safemath{\randmatS}{\amsmathbb{S}}
\safemath{\randmatT}{\amsmathbb{T}}
\safemath{\randmatU}{\amsmathbb{U}}
\safemath{\randmatV}{\amsmathbb{V}}
\safemath{\randmatW}{\amsmathbb{W}}
\safemath{\randmatX}{\amsmathbb{X}}
\safemath{\randmatY}{\amsmathbb{Y}}
\safemath{\randmatZ}{\amsmathbb{Z}}
\safemath{\randmatSigma}{\mathbb{\Sigma}}
\safemath{\randmatPhi}{\mathbb{\Phi}}
\safemath{\randmatLambda}{\mathbb{\Lambda}}

\safemath{\matSigma}{\bm{\Sigma}}
\safemath{\matPhi}{\bm{\Phi}}
\safemath{\matLambda}{\bm{\Lambda}}

\usepackage{glossaries}
\glsdisablehyper
\newglossaryentry{mie}
{
  name={MIE},
  description={maximum index-based estimator},
  first={\glsentrydesc{mie} (\glsentrytext{mie})}
}
\usepackage{pgfplots}
\pgfplotsset{compat=1.14}
\usepgfplotslibrary{groupplots}
\usetikzlibrary{pgfplots.groupplots}

\newtheorem{theorem}{Theorem}
\newtheorem{lemma}{Lemma}
\newtheorem{corollary}{Corollary}

\newtheorem{proposition}{Proposition}

\newcommand{\cce}{\text{\tiny CCE}}
\newcommand{\mmie}{\text{\tiny MMIE}}
\newcommand{\ratedist}{\text{\tiny RD}}
\newcommand{\onebit}{\text{\tiny 1-bit}}

\newcommand {\snr} {\mathtt{SNR}}

\makeatletter
\newcommand\footnoteref[1]{\protected@xdef\@thefnmark{\ref{#1}}\@footnotemark}
\makeatother
\newcommand\barbelow[1]{\stackunder[1.2pt]{$#1$}{\rule{.8ex}{.075ex}}}
\usepackage[lining]{ebgaramond}

\makeatletter
\newsavebox\myboxA
\newsavebox\myboxB
\newlength\mylenA

\newcommand*\mybar[2][0.75]{%
	\sbox{\myboxA}{$\m@th#2$}%
	\setbox\myboxB\null% Phantom box
	\ht\myboxB=\ht\myboxA%
	\dp\myboxB=\dp\myboxA%
	\wd\myboxB=#1\wd\myboxA% Scale phantom
	\sbox\myboxB{$\m@th\overline{\copy\myboxB}$}%  Overlined phantom
	\setlength\mylenA{\the\wd\myboxA}%   calc width diff
	\addtolength\mylenA{-\the\wd\myboxB}%
	\ifdim\wd\myboxB<\wd\myboxA%
	\rlap{\hskip 0.5\mylenA\usebox\myboxB}{\usebox\myboxA}%
	\else
	\hskip -0.5\mylenA\rlap{\usebox\myboxA}{\hskip 0.5\mylenA\usebox\myboxB}%
	\fi}
\makeatother
\makeatletter
\newcommand\customsize{\@setfontsize\customsize{11}{13.6}}
\makeatother
\title{Extremum Encoding for Joint Baseband Signal Compression and Time-Delay Estimation for Distributed Systems}

\name{Amir Weiss$^{\star}$,  Yuval Kochman$^{\dagger}$ and Gregory W. Wornell$^{\ddagger}$}

\address{\customsize
	\begin{tabular}{ccc}
		$^{\star}$Faculty of Engineering & $^\dagger$School of Computer Science and Engineering & $^{\ddagger}$Research Laboratory of Electronics\\
		Bar-Ilan University & The Hebrew University of Jerusalem & Massachusetts Institute of Technology\\
		amir.weiss@biu.ac.il & yuvalko@cs.huji.ac.il  & gww@mit.edu
	\end{tabular}
}

\begin{document}
	\ninept
	\maketitle

	\begin{abstract}
		% In distributed systems, joint processing of observed signals from different sensors is often not possible. Motivated by the proliferation of such systems, we address the fundamental, important time-delay estimation (TDE) problem, but with communication constraints between two non-co-located sensors, where there is a need for a TDE-oriented compression technique. Building on the recently proposed ``extremum encoding"-based compression-estimation scheme, in this work we propose an adapted (/extended) scheme for complex-valued signals, which is suitable for radio-frequency baseband signals. We derive an upper bound on its error probability, and show that it is exponentially tight. Our analytical results are corroborated via simulations,  demonstrating the potential applicability of our scheme to RF systems, and showing its favorable performance relative to two other benchmark methods.
		The ubiquitous time-delay estimation (TDE) problem becomes nontrivial when sensors are non-co-located and communication between them is limited. Building on the recently proposed “extremum encoding” compression-estimation scheme, we address the critical extension to complex-valued signals, suitable for radio-frequency (RF) baseband processing. This extension introduces new challenges, e.g., due to unknown phase of the signal of interest and random phase of the noise, rendering a na\"ive application of the original scheme inapplicable and irrelevant. In the face of these challenges, we propose a judiciously adapted, though natural, extension of the scheme, paving its way to RF applications. While our extension leads to a different statistical analysis, including extremes of non-Gaussian distributions, we show that, ultimately, its asymptotic behavior is akin to the original scheme. We derive an exponentially tight upper bound on its error probability, corroborate our results via simulation experiments, and demonstrate the superior performance compared to two benchmark approaches.
		
	\end{abstract}
	\begin{keywords}
		Time-delay estimation, data compression, distributed estimation, compression for estimation, max-index estimator.
	\end{keywords}
	\vspace{-0.3cm}
	\section{Introduction}\label{sec:intro}
	\vspace{-0.2cm}
	Time-delay estimation (TDE) remains a fundamental problem at the core of numerous applications in different physical domains, such as acoustics and radio-frequency (RF). These applications span a wide range of areas, with localization and tracking being central among them \cite{musicki2009mobile,yeredor2010joint,weiss2022semi}. Given its critical importance, TDE has been extensively studied over the past decades, resulting in a rich body of literature \cite{ziv1969some,quazi1981overview,ianniello1982time,weiss1983fundamental,weinstein1984fundamental,azaria1984time,fertner1986comparison,carter1987coherence,jacovitti1993discrete,brandstein1997robust,bjorklund2003survey,benesty2004time,chen2006time}.
	
	However, building on recent technological advancements, giving rise to new concepts such as the internet-of-things \cite{da2014internet}, there is an increasing need to address communication constraints (possibly through data compression) in various distributed estimation tasks, and specifically in TDE as well. For this reason, the standard assumption that the central computing unit has access (in the fundamental case of two sensors) to both of the raw received signals may no longer hold for some of the modern (and future) applications. For instance, envision a collection of individual sensors, which (by design) are low-cost and have limited resources in terms of energy and their communication abilities. Upon the demand of some holistic central computing unit, they become an \emph{ad-hoc} distributed system for the sake of a specific task, such as localization of an entity present in their relevant sensing range.
	
	In light of such potential scenarios, it is essential to minimize the resource requirements of the spatially distributed (i.e., non-co-located) low-cost devices. This is similar to the case of sensor networks \cite{akyildiz2002survey,chen2002source}, which consist of numerous small, low-power devices that operate for a collective inference task (e.g., \cite{wang2003preprocessing}). Of course, the proliferation of RF wireless devices sets this domain as a particularly important one.
	
	The motivation above has led to a renewed interest in compression for TDE, e.g., \cite{vasudevan2003application,fowler2005fisher,chen2010data,fuyong2012data,vargas2018compressed}. However, it turns out that not more than a handful of works consider the fact that a \emph{joint} compression-estimation scheme can lead to a potentially different, more efficient signal compression techniques, rather than applying standard signal compression methods to the received signals. Specifically, in our recent work \cite{weiss2024joint}, we proposed a joint compression-estimation method for TDE with communication constraints for a real-valued signal model, focusing on theoretical aspects and establishing key preliminary results. Extending this foundation, we now consider the complex-valued signal model, which is better suited for RF baseband signals. Hence, this work expands the applicability of our approach to this important, prevalent domain.
	
	We consider a discrete-time formulation of the TDE problem for two non-co-located sensors, where joint processing of the two raw baseband signals is not possible due to communication constraints. Following \cite{weiss2024joint}, we continue to focus on the fundamental theoretical aspects of the problem, and establish the critical extension to this widespread setting, which paves the way to additional important practical extensions.
	
	Building on \cite{weiss2024joint}, our main contributions in this paper are as follows:
	\begin{itemize}[itemsep=0.15pt]
		\item \emph{A joint compression and TDE method for complex-valued baseband signals}: We propose an extremum-encoding-based compression and an accompanying time-delay estimator that can be applied to complex-valued signals, thereby broadening the scope and applicability of our approach to (possibly RF) baseband signals.
		\item \emph{Performance Analysis}: We derive an upper bound on the error probability of the proposed scheme, which decays exponentially with the number of bits sent from one receiver to the other. This shows that our estimator is consistent in the communication sense. We further show that the exponential decays of our bound is (exponentially) tight for our scheme. Our findings are corroborated through empirical simulation results.
	\end{itemize}
	
	\vspace{-0.3cm}
	\section{Problem Formulation}\label{sec:problemformulation}
	\vspace{-0.2cm}
	We start by formulating a simplified version of the TDE problem. In particular, consider the observed discrete-time signals at two distant sensors, 
	\begin{equation}\label{eq:initialmodel}
		\begin{aligned}
			\rndr_1[n] &= \rnds[n] + \rndz_1[n]\in\complexset, \;\; &\text{(sensor 1)} \\
			\rndr_2[n] &= \rnds[n-\rndd]\alpha e^{\jmath\widetilde{\theta}} + \rndz_2[n]\in\complexset, \;\; &\text{(sensor 2)}
		\end{aligned}
	\end{equation}
	where
	\begin{itemize}
		\item $\rnds[n]\overset{\text{iid}}{\sim}\jpg(0,1)$ is circular complex normal (CCN), and is the signal that is observed by both sensors with a relative time-delay $\rndd\in\setD$, where $\setD\triangleq\{-d_m,\ldots,d_m\}$ is the ``delay spread" and $d_m\in\naturals$ is the maximum (absolute) delay;
		\item $e^{\jmath\widetilde{\theta}}=\cos(\widetilde{\theta})+\jmath\sin(\widetilde{\theta})$ is a phase rotation, where $\widetilde{\theta}\in[0,2\pi)$, and $\alpha\in\positivereals$ is an attenuation coefficient, both of which are deterministic and unknown; and
		\item $\rndz_1[n]\overset{\text{iid}}{\sim} \jpg(0,\sigma_1^2), \rndz_2[n]\overset{\text{iid}}{\sim} \jpg(0,\sigma_2^2)$ are statistically independent (white) CCN processes with unknown deterministic variances $\sigma_1^2, \sigma_2^2$, which are also statistically independent of $\rnds[n]$.
	\end{itemize}
	
	\begin{figure}
		\centering
		\includegraphics[width=0.9\columnwidth]{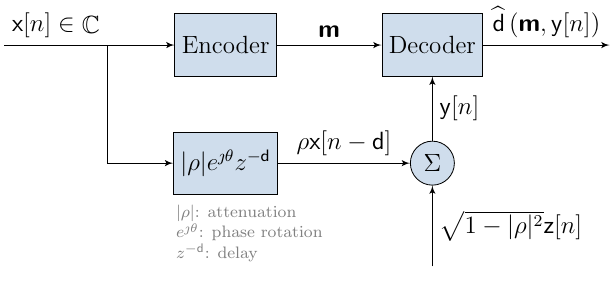}\vspace{-0.3cm}
		\caption{Illustration of the distributed time-delay estimation problem considered in this work. The encoder observes $\rndx[n]$ and generates a message of length $k$ bits $\rvecm\in\{0,1\}^{k\times 1}$. The decoder observes $\rndy[n]$ and receives $\rvecm$, from which it constructs $\widehat{\rndd}(\rvecm,\rndy[n])$, an estimator of $\rndd$.}
		\label{fig:blockdiagram}\vspace{-0.4cm}
	\end{figure}
	For ease of notation, we begin with the following proposition that allows us to continue the analysis with a simplification of the model.\vspace{-0.2cm}
	\begin{proposition}\label{proposition1}
		Model \eqref{eq:initialmodel} is statistically equivalent to the model
		\begin{equation}\label{eq:model}
			\begin{aligned}
				&\rndx[n]\in\complexset, &\text{\emph{(sensor 1, ``encoder")}} \\
				&\rndy[n] = \rho\rndx[n-\rndd] + \bar{\rho}\rndz[n]\in\complexset, &\text{\emph{(sensor 2, ``decoder")}}
			\end{aligned}
		\end{equation}
		depicted in Fig.~\ref{fig:blockdiagram}, where $\rndx[n],\rndz[n]\overset{\text{\emph{iid}}}{\sim} \jpg(0,1)$ are statistically independent, $\rho\triangleq|\rho| e^{\jmath\theta}$ is the correlation coefficient between $\rndx[n]$ and $\rndy^*[n+\rndd]$, where $|\rho|\in(0,1]$ and $\theta\in[0,2\pi)$, such that $|\rho|$ is related to the signal-to-noise ratios (SNRs) of \eqref{eq:initialmodel}, i.e., to $1/\sigma_1^2$ and $1/\sigma_2^2$, and $\bar{\rho}\triangleq\sqrt{1-|\rho|^2}$.
	\end{proposition}
	The (simple) proof is omitted due to space considerations, but it is easy to see that, up to power (/variance) normalization of the observed signals, the processes in both pairs $\rndr_1[n],\rndr_2[n]$ and $\rndx[n],\rndy[n]$ are each marginally white CCN, and are also jointly complex normal with an identical cross-correlation.
	Therefore, we will henceforth work with model \eqref{eq:model}, and accordingly, we define $\snr\triangleq\frac{|\rho|^2}{|\bar{\rho}|^2}$ as the SNR (with $|\rho|^2=\frac{1}{1+\snr^{-1}}$).\footnote{Observe that $\snr\xrightarrow[]{|\rho|\to1}\infty$ and $\snr\xrightarrow[]{|\rho|\to0}0$, as desired.}% For the exact relation between $\snr$ and $1/\sigma_1^2,1/\sigma_2^2$ (SNRs in model \eqref{eq:initialmodel}).
	
	In \eqref{eq:model}, one sensor (the ``encoder") observes the signal $\rndx[n]$ and needs to produce a message $\rvecm\in\{0,1\}^{k\times 1}$ of length $k\in\naturals$ bits to be sent to the other sensor (the ``decoder"). The latter observes $\rndy[n]$, a noisy version of (the $\rho$-scaled) $\rndx[n]$, delayed by $\rndd$ samples. We consider a Bayesian setting in which $\rndd\sim\mathcal{U}(\setD)$ (i.e., uniformly distributed), and the goal of sensor 2, which is assumed to be located at the central computing unit, is to estimate $\rndd$ based on the observed signal $\rndy[n]$ and the received message $\rvecm$, so as to minimize the risk $\Exop\left[\ell(\widehat{\rndd},\rndd)\right]$ for a loss function $\ell:\setD\times\setD\to\positivereals$, where the expectation is with respect to all sources of randomness, i.e., $\rndx[n],\rndz[n]$ and $\rndd$. In this work, we focus on the error loss $\ell(a,b)=\mathbbm{1}_{a\neq b}$ that yields the error probability risk. We are interested in the trade-off between the number of bits $k$ and the error probability.
	\vspace{-0.15cm}
	\section{A Joint Compression-Estimation Scheme}\label{sec:compressestimate}
	\vspace{-0.15cm}
	Our proposed scheme---a natural extension of \cite{weiss2024joint}---is as follows. 
	
	\textbf{Extremum Encoding:} The encoder observes a sequence of length $N=2^{k}$,\footnote{This assumption is for notational convenience, and can be relaxed to $N\in\naturals$.} namely $\mathcal{X}_N\triangleq\{\rndx[n]\}_{n=0}^{N-1}$, and sends as the message $\rvecm$ (of length $k$ bits) the \emph{index} of the maximum magnitude among all $\mathcal{X}_N$,
	{\setlength{\belowdisplayskip}{4pt} \setlength{\belowdisplayshortskip}{4pt}
		\setlength{\abovedisplayskip}{4pt} \setlength{\abovedisplayshortskip}{4pt}
		\begin{equation}\label{definitionofJ}
			\rndj\triangleq \arg \max_{0\leq n\leq N-1} \; \left|\rndx[n]\right|^2,% \; \Longrightarrow \; \rvecm= {\rm{dec\_to\_bin}}\left(\rndj\right),
	\end{equation}}
	where $\rvecm\in\{0,1\}^{k\times 1}$ is the binary representation of $\rndj$.
	
	\textbf{Maximum-Magnitude-Index-Based Estimation:} The decoder, which in particular observes $\{\rndy[n]\}_{n=-d_m}^{N-1+d_m}$, upon receiving the message $\rvecm$ (equivalently, the index $\rndj$), constructs% the following estimator,
	{\setlength{\belowdisplayskip}{4pt} \setlength{\belowdisplayshortskip}{4pt}
		\setlength{\abovedisplayskip}{4pt} \setlength{\abovedisplayshortskip}{4pt}
		\begin{equation}\label{eq:maxestsimple}
			{\widehat{\rndd}}_{\mmie} \triangleq \arg\max_{\ell\in\setD} \; \left|\rndy[\rndj+\ell]\right|^2,
	\end{equation}}
	which we call the ``maximum-magnitude-index"-based estimator (MMIE). Differently from \cite{weiss2024joint}, this scheme is designed to a complex-valued random signal, which generally leads to different statistical characteristics. Still, the proposed scheme maintains its intuitive interpretation: the encoder sends as the message the center of the decoder's search interval, whose size is the delay spread, $|\setD|$. Then, the chosen estimated time-delay is the shift (in the opposite direction, c.f.\ \eqref{eq:model}) relative to that center, for which the magnitude of the noisy signal at the decoder is maximized.
	
	\vspace{-0.3cm}
	\subsection{A Cross Correlation-Based Interpretation of Our Scheme}\label{subsec:interpretation}
	The underlying logic of \eqref{eq:maxestsimple} is, satisfactorily, quite intuitive. To see it more clearly, let us first consider the simple, commonly used cross-correlation estimator (CCE) of $\rndd$ from $\rndx[n]$ and $\rndy[n]$, given by\footnote{We emphasize that here, as well as later for the MMIE, we use $\left|\widehat{\rho}_{\cce}(\ell)\right|^2$ rather than $\widehat{\left|\rho_{\cce}(\ell)\right|^2}$, an abuse of notation, for brevity. We underscore that this is \emph{not} the squared-absolute value of an estimator $\widehat{\rho}_{\cce}(\ell)$. Rather, our one intention is for the estimator of the squared-absolute value of the correlation coefficient.}
	\begin{equation}\label{eq:ccedefinition}
		\widehat{\rndd}_{\cce} = \arg\max_{\ell\in\setD} \; \frac{1}{N}\left|\sum_{n=0}^{N-1}\rndx[n]\rndy^*[n+\ell]\right|^2 \triangleq \arg\max_{\ell\in\setD} \; \left|\widehat{\rho}_{\cce}(\ell)\right|^2.
	\end{equation}
	Interestingly, it can be shown that the MMIE \eqref{eq:maxestsimple} is mimicking the CCE \eqref{eq:ccedefinition}, designed for a constraints-free communication setting, which is simply trying to identify the time-lag at which the cross-correlation between $\rndx[n]$ and $\rndy[n]$ is maximized. To see this, consider first the following result.
	
	\begin{lemma}\label{lemma1}
		Consider model \eqref{eq:model} with $\rndd\equiv0$, namely, $\{\rndx[n]\}$ and $\{\rndy[n]\}$ are zero-mean unit-variance white (i.e., uncorrelated) CCN processes with a complex-valued correlation coefficient $\rho$. With $\rndj$ as in \eqref{definitionofJ}, let 
		{\setlength{\belowdisplayskip}{4pt} \setlength{\belowdisplayshortskip}{4pt}
			\setlength{\abovedisplayskip}{4pt} \setlength{\abovedisplayshortskip}{4pt}
			\begin{equation}
				|\widehat{\rho}_{\emph{\mmie}}|^2 \triangleq \frac{|\rndy[\rndj]|^2}{\Exop[|\rndx[\rndj]|^2]}.
		\end{equation}}
		Then, $|\widehat{\rho}_{\emph{\mmie}}|^2$ is an asymptotically unbiased estimator of $|\rho|^2$ with
		\begin{equation}\label{eq:varofcorrelationmaxestimator}
			\Varop\left(|{\widehat{\rho}}_{\emph{\mmie}}|^2\right) = \setO\left(\tfrac{1}{H(N)}\right)=\setO\left(\tfrac{1}{k}\right),
		\end{equation}
		where $H(N)\triangleq\sum_{n=1}^N\tfrac{1}{n}=\log(N)+\gamma+\setO(\tfrac{1}{N})$
		% \begin{equation}
			% H(N)\triangleq\sum_{n=1}^N\tfrac{1}{n}=\log(N)+\gamma+\setO(\tfrac{1}{N})= k\log(2)(1+o(1)),    
			% \end{equation}
		is the harmonic series and $\gamma\approx0.577$ is the Euler–Mascheroni constant.
	\end{lemma}\vspace{-0.2cm}
	\textit{Proof Sketch:} The proof is based on the fact that $|\rndx[n]|^2$ is exponentially distributed with a parameter rate $1$, which leads to the distribution $\prob{|\rndx[\rndj]|^2-\log(N)<x}=\left(1-\frac{e^{-x}}{N}\right)^N\xrightarrow[N\to\infty]{} e^{-e^{-x}}$. Loosely speaking, this means that $|\rndx[\rndj]|^2$ is asymptotically approximately deterministic relative to the noise $\{\rndz[n]\}$. A (rather technical) analysis yields
	{\setlength{\belowdisplayskip}{4pt} \setlength{\belowdisplayshortskip}{4pt}
		\setlength{\abovedisplayskip}{4pt} \setlength{\abovedisplayshortskip}{4pt}
		\begin{align}
			\hspace{-0.2cm}&\Exop\left[|\rndx[\rndj]|^2\right]\hspace{-0.05cm}=\hspace{-0.05cm}H(N),\, \Varop\left(|\rndx[\rndj]|^2\right)\hspace{-0.05cm}=\hspace{-0.05cm}\sum_{n=1}^N\frac{1}{n^2}\hspace{-0.05cm}=\hspace{-0.05cm}\frac{\pi^2}{6}(1+o(1)),\\
			\hspace{-0.2cm}&\Exop\left[|\rndy[\rndj]|^2\right]\hspace{-0.05cm}=\hspace{-0.05cm}|\rho|^2H(N)(1+o(1)), \Varop\left(|\rndy[\rndj]|^2\right)\hspace{-0.05cm}=\hspace{-0.05cm}\setO\left(H(N)\right)\\
			\hspace{-0.2cm}&\Rightarrow \Exop\left[|{\widehat{\rho}}_{\mmie}|^2\right]=|\rho|^2(1+o(1)), \Varop\left(|{\widehat{\rho}}_{\mmie}|^2\right)=\setO\left.(\tfrac{1}{H(N)}\right).
	\end{align}}
	\hfill $\blacksquare$
	
	% \noindent The detailed, full proof of of Lemma \ref{lemma1} is available at \cite{weiss2025icasspsuppmat}.
	
	An immediate consequence of Lemma \ref{lemma1} is that $|{\widehat{\rho}}_{\mmie}|^2$ converges in mean-square to $|\rho|^2$, and therefore also in probability.
	
	This leads us to the natural definition
	\begin{equation}\label{eq:correlationmaxestimator}
		|\widehat{\rho}_{\mmie}(\ell)|^2 \triangleq \frac{|\rndy[\rndj+\ell]|^2}{\Exop\left[|\rndx[\rndj]|^2\right]}, \quad \forall \ell\in\integers,
	\end{equation}
	which is, of course, an asymptotically unbiased, consistent estimator of $|\rho(\ell)|^2\triangleq\left|\Exop\left[\rndy[n]\rndx[n-\ell]|\rndd\right]\right|^2$. This is due to the fact that for any time shift $\ell$, we are again at the same setting considered in Lemma \ref{lemma1}, but for a different value of the correlation coefficient. We emphasize that, although \eqref{eq:correlationmaxestimator} is strictly positive with probability $1$, this also holds true for the (possibly slightly counter-intuitive) case $\rho=0$.
	
	With \eqref{eq:correlationmaxestimator}, we can revisit \eqref{eq:maxestsimple}, and since $\Exop[|\rndx[\rndj]|^2]$ is independent of the optimization index $\ell$, we have,
	{\setlength{\belowdisplayskip}{5pt} \setlength{\belowdisplayshortskip}{5pt}
		\setlength{\abovedisplayskip}{5pt} \setlength{\abovedisplayshortskip}{5pt}
		\begin{equation}\label{eq:maxestsimpleascorr}
			{\widehat{\rndd}}_{\mmie} = \arg\max_{\ell\in\setD} \; \frac{|\rndy[\rndj+\ell]|^2}{\Exop\left[|\rndx[\rndj]|^2\right]}
			=\arg\max_{\ell\in\setD} \; \left|\widehat{\rho}_{\mmie}(\ell)\right|^2.
	\end{equation}}
	As easily seen, the MMIE \eqref{eq:maxestsimpleascorr} (for constrained communication) and the CCE \eqref{eq:ccedefinition} (for constraints-free communication) are akin---both choose the time-lags at which their respective estimated squared-magnitudes of the cross-correlation are maximized as the estimated delays.
	\vspace{-0.25cm}
	% \subsection{Computational Complexity}\label{subsec:somplexity}
	\subsection{Algorithmic Complexity}\label{subsec:somplexity}
	\vspace{-0.15cm}It should not come as a surprise that our proposed, adapted encoding-estimation scheme enjoys the same computational complexity as the originally proposed scheme \cite{weiss2024joint} for the slightly more simple signal model. However, and before we rigorously address the statistical performance of the MMIE in Section \ref{sec:asymptoticperformance} that follows, for completeness of the exposition, we briefly explain here its merit in terms of its complexity. As a benchmark, recall that for $N=2^k$ samples, computing a standard cross-correlation-based estimator at $|\setD|=2d_m+1$ time-lags is $\setO(d_m2^k)$. For the MMIE \eqref{eq:maxestsimple}, one simply needs to perform two searches for the maxima of two arrays of size $2^k$ and $|\setD|$ (encoder and decoder, respectively). Consequently, the overall complexity of our scheme is $\setO(2^k+d_m)$. Thus, our proposed method, on top of being communication-efficient, is more computationally economic.
	% \vspace{-0.15cm}Beyond its performance and statistical consistency (discussed in Section \ref{sec:asymptoticperformance} below), the MMIE \eqref{eq:maxestsimple} also enjoys reduced computational complexity, specifically relative to standard cross-correlation-based estimators, such as the CCE \eqref{eq:ccedefinition}. For such standard cross-correlators, computing the empirical cross-correlation at $|\setD|=2d_m+1$ time-lags based on $2^k$ samples amounts to $\setO(d_m2^k)$ operations. In contrast, the MMIE simply requires two array searches for the maximum of two arrays of sizes $2^k$ (encoder) and $|\setD|$ (decoder), hence its complexity is $\setO(2^k+d_m)$. Thus, our proposed method is not only more communication-efficient, but is also attractive in terms of the computational resources it requires.
	\vspace{-0.2cm}
	% \section{Performance Analysis of the MMIE}\label{sec:asymptoticperformance}
	\section{Statistical Performance Analysis}\label{sec:asymptoticperformance}
	\vspace{-0.15cm}
	While the intuition of \eqref{eq:maxestsimple} provided above is encouraging, it is still necessary to provide an accompanying analysis for performance guarantees. It turns out that due to the asymptotic concentration of $|\rndx[\rndj]|^2$---an asymptotically Gumble \cite{david2004order} random variable (RV)---around its mean, the intuition behind \eqref{eq:maxestsimple} discussed above can be rigorously verified. Concretely, we have the following result, whose proof appears in Section \ref{subsec:proofs}.
	\begin{theorem}[Error probability upper and lower bounds]\label{theorem1}
		Consider the error loss $\ell(a,b)=\mathbbm{1}_{a\neq b}$, which yields the error probability risk $\epsilon\triangleq\prob{\widehat{\rndd}_{\emph{\mmie}}\neq \rndd}$. Then, in the asymptotic regime $k\to\infty$ and for any $|\rho|\in(0,1]$,\footnote{The case $\rho=0$ is less interesting, and rather trivial to analyze, since both sensors observe (purely) statistically independent white CCN processes.} we have,
		\begin{align}
			\barbelow{\epsilon}(k,\rho)\left(1+o(1)\right) \leq \epsilon \leq \bar{\epsilon}(k,\rho,d_m)\left(1+o(1)\right)\label{eq:upperandlowerbound},
		\end{align}
		where, denoting $E(\snr)\triangleq\frac{\snr}{2+\snr}=\frac{|\rho|^2}{2-|\rho|^2}$,
		\begin{align}
			\bar{\epsilon}(k,\snr,d_m) &\triangleq \tfrac{2d_m(1+\snr)}{2+\snr}\cdot 2^{-k\cdot E(\snr)(1-\varepsilon(k))}, \label{eq:upperbound}\\
			\barbelow{\epsilon}(k,\snr) &\triangleq 2^{-k\cdot E({\snr})(1+\varepsilon(k))}\label{eq:lowerbound},
		\end{align}
		where $\varepsilon(k)=o(1)$ and the $o(1)$ terms go to zero as $k\to\infty$.
	\end{theorem}
	\vspace{-0.15cm}
	An immediate corollary of Theorem \ref{theorem1} is the following.\vspace{-0.15cm}
	\begin{corollary}[``Communication Consistency"]\label{corollary1}
		For the same setting of Theorem \ref{theorem1}, the MMIE \eqref{eq:maxestsimple} is consistent in the communication sense, namely with respect to the message size (number of transmitted bits), such that,
		\begin{equation}\label{eq:consistencyofmie}
			\lim_{k\to\infty}\prob{\widehat{\rndd}_{\emph{\mmie}}\neq \rndd}=0.
		\end{equation}
		Moreover, $d_m$ need \emph{not} be fixed, and it is only required that $d_m=o\left(2^{k\cdot E(\snr)}\right)$. In other words, the delay spread can grow with the observation time, as long as it grows ``sufficiently slow" with $k$. Additionally, we note in passing that $\widehat{\rndd}_{\emph{\mmie}}$ is also consistent in the classical sense, with respect to the encoder's observation interval, $N$, i.e., $\prob{\widehat{\rndd}_{\emph{\mmie}}\neq \rndd}\xrightarrow[]{N\to\infty}0$.
	\end{corollary}
	
	Another corollary of Theorem \ref{theorem1}, which establishes the asymptotic error probability rate of decay of the MMIE, is the following.
	
	\begin{corollary}[Error exponent]\label{corollary2}
		In the setting of Theorem \ref{theorem1},
		\begin{equation}\label{eq:exactasymptoticerrorexponent}
			\lim_{k\to\infty}-\frac{1}{k}\log_2(\epsilon) = E\left(\snr\right) = \frac{\snr}{2+\snr}.
		\end{equation}
	\end{corollary}
	\noindent\textbf{Proof of Corollary \ref{corollary2}} Sandwich $-\frac{1}{k}\log_2(\epsilon)$ with $-\frac{1}{k}\log_2\left(\barbelow{\epsilon}(k,\rho)\right)$ and $-\frac{1}{k}\log_2(\bar{\epsilon}(k,\rho,d_m))$ from above and below, respectively, and take the limit $k\to\infty$ to obtain \eqref{eq:exactasymptoticerrorexponent}. \hfill $\blacksquare$
	
	Interestingly, we observe that $\rho=1$ (the ``infinite SNR regime") is insufficient for $\epsilon$ to be zero. Indeed, one of the samples at the ``edges" of the time-interval that are observed by the decoder---but not by the encoder (due to the time-delay uncertainty)---can be greater than the one observed by the encoder. Hence an error may occur, regardless of the SNR, due to an ``unlucky" realization of the time-delay $\rndd$ and $\rndj$, the location of the maximum within $\{1,\ldots,N\}$.
	
	\vspace{-0.2cm}
	\subsection{Proof of Theorem \ref{theorem1}}\label{subsec:proofs}
	To establish the upper and lower bounds \eqref{eq:upperbound} and \eqref{eq:lowerbound}, respectively, we shall use the following two lemmas.
	\begin{lemma}\label{lemma2}
		Let $\rndj\in\{1,\ldots,N\}$ as defined in \eqref{definitionofJ}. Then,
		\begin{equation}
			p(\tau)\triangleq\prob{|\rndx[\rndj]|^2<\log(N)+\tau}=\left(1-\tfrac{e^{-\tau}}{N}\right)^N\leq e^{-e^{-\tau}}.
		\end{equation}
		In particular, let $\delta(k)\triangleq\log(\log(N)))=\log(k\log(2))$, with which
		\begin{equation}
			p(-\delta(k))\leq 2^{-k}, \; p(\delta(k))=1-o(1).\label{eq:usefulprobbounds}
		\end{equation}
		
	\end{lemma}
	\begin{lemma}\label{lemma3}
		Let $\rndv,\rndz\sim\jpg(0,1)$ be independent, and $\rndu\triangleq \min(\rndv,V)\tfrac{\rndv}{|\rndv|}$, for some $V>\tfrac{1}{\sqrt{2}}$. Then, for any $a\in\positivereals$, (recall that $|\rho|^2+\bar{\rho}^2=1$)
		\begin{equation}\label{eq:exponentialapporxequality}
			\prob{|\rho\rndu+\bar{\rho}\rndz|>a} \geq \prob{|\rndv|>a} - 2Ve^{-V^2}.
		\end{equation}
	\end{lemma}\vspace{-0.2cm}
	\noindent\textbf{Proof of Lemma \ref{lemma2}} Since $\rndx[n]\sim\jpg(0,1)$ is iid,
	\begin{align}
		&\prob{|\rndx[\rndj]|^2<\log(N)+\tau}\overset{\text{(a1)}}{=}\left(1-\tfrac{e^{-\tau}}{N}\right)^N\overset{\text{(b1)}}{\leq} e^{-e^{-\tau}},
	\end{align}
	where we have used in (a1) that $|\rndx[n]|^2$ is exponentially distributed with a rate $1$ and the independence of $\{\rndx[n]\}_{n=1}^N$; and in (b1) the inequality $(1-x)^N\leq e^{-Nx}$. An application of the above yields \eqref{eq:usefulprobbounds}. \hfill $\blacksquare$
	
	The proof of Lemma \ref{lemma3} is rather technical and straightforward, and is therefore omitted due to space considerations. We note, however, that when $V$ increases, the magnitude-truncation of $\rndu$ becomes (exponentially) negligible, so we expect the probabilities on both sides of \eqref{eq:exponentialapporxequality} to approximately coincide. Indeed, this happens (exactly) when $V\to\infty$.
	
	We now prove Theorem \ref{theorem1}. For the upper bound, let $\rndv\sim\jpg(0,1)$ and denote the index $\ell\neq\rndd$, with which we have,
	\begin{align}
		&\prob{\widehat{\rndd}\neq \rndd\mid \rndx[\rndj],\rndd}\\
		&\overset{\text{(a2)}}{\leq}2d_m\prob{|\rndy[\rndj+\ell]|>|\rndy[\rndj+\rndd]|\mid \rndx[\rndj],\rndd}\\
		&\overset{\text{(b2)}}{\leq}2d_m\Exop\left[\left.\prob{|\rndv|>|\rndy[\rndj+\rndd]|\mid \rndx[\rndj],\rndd,\rndz[\rndj+\rndd]}\right|\rndx[\rndj],\rndd\right] \\ 
		&\overset{\text{(c2)}}{=} \Exop\left[\left.e^{-|\rndy[\rndj+\rndd]|^2}\right|\rndx[\rndj],\rndd\right]\overset{\text{(d2)}}{=}\Exop\left[\left.e^{-\frac{\bar{\rho}^2}{2}|\sqrt{2}\frac{\rho}{\bar{\rho}}\rndx[\rndj]+\rndz[\rndj+\rndd]|^2}\right|\rndx[\rndj]\right] \label{eq:startequalitiesthatweusetwice}\\
		&\overset{\text{(e2)}}{=}\tfrac{2d_m}{2-|\rho|^2}e^{-|\rndx[\rndj]|^2\frac{|\rho|^2}{2-|\rho|^2}}\label{eq:mgfofnoncentralchisquare},
	\end{align}
	where we have used in (a2) the union bound; in (b2) the law of total expectation (LTE) and we have replaced $\rndx[\rndj+\ell-\rndd]$ with a standard CCN RV (and in particular, whose magnitude is not bounded, thus increasing the integration domain); in (c2) the fact that $|\rndv|$ is Rayleigh distributed with a scale parameter $\frac{1}{2}$; in (d2) the fact that $\rndz[\rndj+\rndd]$ is independent of $\rndd$ (since it is a white CCN noise); and in (e2) the moment generating function of a noncentral chi-square RV with two degrees of freedom and a noncentrality parameter $2\tfrac{|\rho|^2}{\bar{\rho}^2}|\rndx[\rndj]|^2=2\snr|\rndx[\rndj]|^2$.
	
	Now, simply apply the LTE with Lemma \ref{lemma2} to have,
	\begin{align}
		\prob{\widehat{\rndd}\neq \rndd} &= \Exop\left[\prob{\widehat{\rndd}\neq \rndd\mid \rndx[\rndj],\rndd}\right]\\
		&\leq p(-\delta(k)) + \tfrac{2d_m}{2-|\rho|^2}e^{-\log(N)(1-\varepsilon(k))\frac{|\rho|^2}{2-|\rho|^2}}\\
		&=\tfrac{2d_m}{2-|\rho|^2}2^{-k\frac{|\rho|^2}{2-|\rho|^2}(1-\varepsilon(k))}(1+o(1)),\label{eq:lasttransitionupperbpund0}
	\end{align}
	where we have used \eqref{eq:usefulprobbounds} in \eqref{eq:lasttransitionupperbpund0}, thus establishing the upper bound \eqref{eq:upperbound}.
	
	As for the lower bound \eqref{eq:lowerbound}, it is easy to see that, for any $\ell\in\setD\backslash\{\rndd\}$,
	\begin{align}\label{eq:sinplelowerbound}
		&\prob{\widehat{\rndd}\neq \rndd\mid \rndx[\rndj],\rndd}\\
		&\overset{\text{(a3)}}{\geq}\Exop\left[\left.\prob{|\rndy[\rndj+\ell]|>|\rndy[\rndj+\rndd]|\mid \rndx[\rndj],\rndd,\rndz[\rndj+\rndd]}\right|\rndx[\rndj],\rndd\right],\\
		&\overset{\text{(b3)}}{\geq}\Exop\left[\left.\prob{|\rndv|>|\rndy[\rndj+\rndd]|\mid \rndx[\rndj],\rndd,\rndz[\rndj+\rndd]}\right|\rndx[\rndj],\rndd\right]-2|\rndx[\rndj]|e^{-|\rndx[\rndj]|^2},\\
		&\overset{\text{(c3)}}{=}\tfrac{1}{2-|\rho|^2}e^{-|\rndx[\rndj]|^2\frac{|\rho|^2}{2-|\rho|^2}}-2|\rndx[\rndj]|e^{-|\rndx[\rndj]|^2}
	\end{align}
	where we have used in (a3) the LTE and taking only one event from the union of (statistically equivalent) $2d_m$ events; in (b3) Lemma \ref{lemma3}; and in (c3) the same chain of equalities as in \eqref{eq:startequalitiesthatweusetwice}--\eqref{eq:mgfofnoncentralchisquare}. Now, we again use the LTE with Lemma \ref{lemma2} to obtain,
	\begin{align}
		&\prob{\widehat{\rndd}\neq \rndd}\\
		&\geq \tfrac{1}{2-|\rho|^2}\Exop\left[e^{-|\rndx[\rndj]|^2\frac{|\rho|^2}{2-|\rho|^2}}\right] - 2\Exop\left[|\rndx[\rndj]|e^{-|\rndx[\rndj]|^2}\right] \\
		&\geq\tfrac{1}{2-|\rho|^2}2^{-k\frac{|\rho|^2}{2-|\rho|^2}(1+\varepsilon(k))}p(\delta(k))-\sqrt{2}e^{-\tfrac{1}{2}}p(-\delta(k)),\label{eq:lasttransitionupperbpund}\\
		&=\tfrac{1}{2-|\rho|^2}2^{-k\frac{|\rho|^2}{2-|\rho|^2}(1+\varepsilon(k))}(1+o(1)),
	\end{align}
	where we have used $2|t|e^{-t^2}\leq\sqrt{2}e^{-\frac{1}{2}}$ for any $t\in\reals$ in \eqref{eq:lasttransitionupperbpund}, thus establishing \eqref{eq:lowerbound}. Finally, we the use the relation $|\rho|^2=\frac{1}{1+\snr^{-1}}$ to have $\frac{|\rho|^2}{2-|\rho|^2}=\frac{\snr}{2+\snr}$, which completes the proof of the theorem. \hfill $\blacksquare$
	
	% \noindent\textbf{Proof of Theorem \ref{theorem3}} Sandwiching $-\frac{1}{k}\log_2(\epsilon)$ of the left-hand side of \eqref{eq:exactasymptoticerrorexponent} with $-\frac{1}{k}\log_2(\bar{\epsilon}(N,\rho,d_m))$ and $-\frac{1}{k}\log_2\left(\barbelow{\epsilon}(N,\rho)\right)$ from below and above, respectively, and taking $k\to\infty$ proves the theorem. \hfill $\blacksquare$
	\vspace{-0.4cm}
	\section{Simulation Results}\label{sec:simulationresults}
	\vspace{-0.2cm}
	We now present results of a simulation experiment that corroborates our analysis and demonstrates that our method outperforms: (i) a rate-distortion (RD) signal compression benchmark and; (ii) the ubiquitous $1$-bit per sample scalar quantization approach (e.g., \cite{weiss2021one,zhang2021direct,ni2023detection}).
	
\begin{figure}[t!]
		\centering
		\begin{subfigure}[t]{0.485\columnwidth}
			\centering
			\includegraphics[width=\columnwidth, trim=2cm 0cm 18cm 0.75cm, clip]{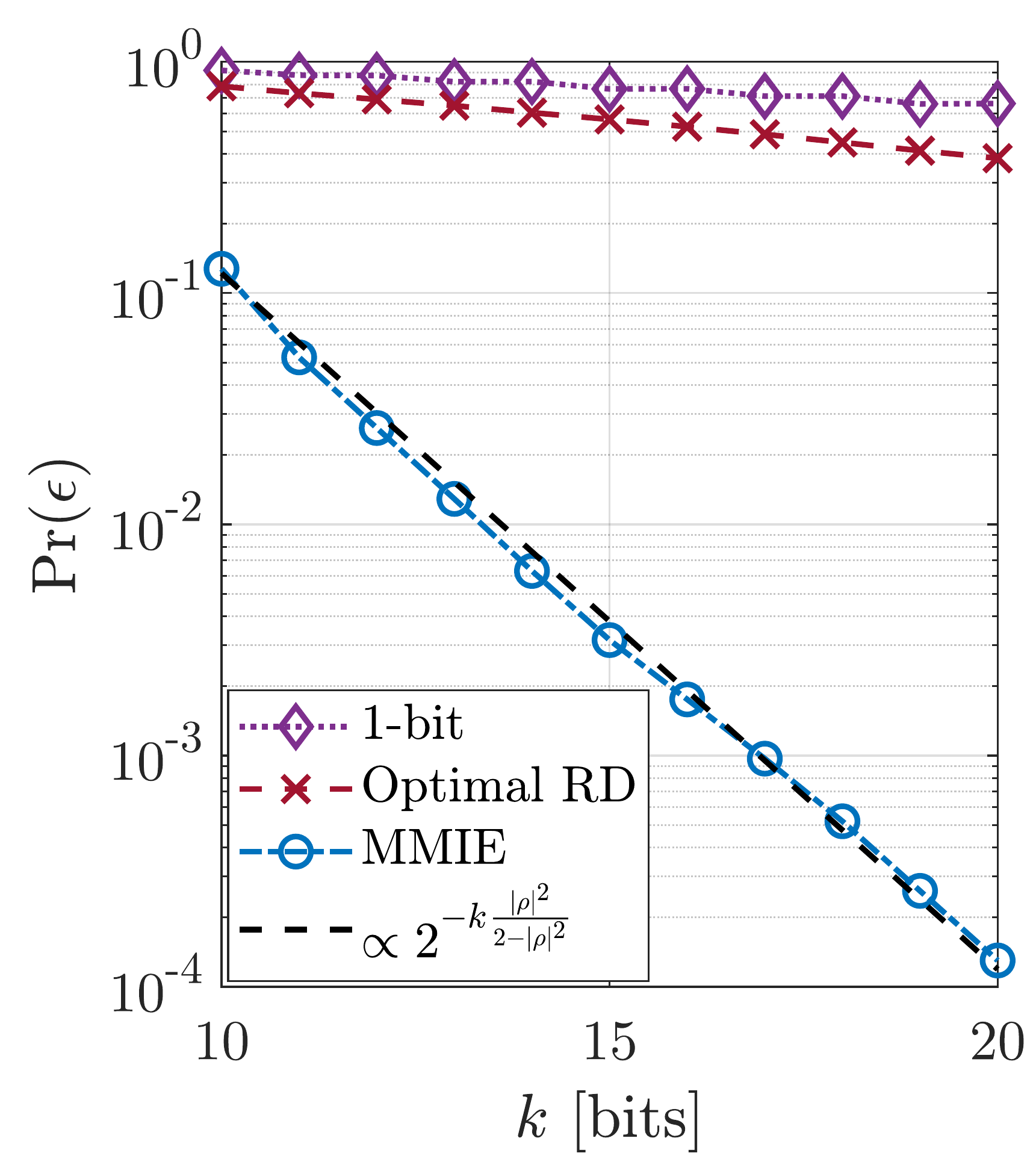}
			\caption{}
			\label{fig:subfig1}
		\end{subfigure}
		% \hfill
		\begin{subfigure}[t]{0.485\columnwidth}
			\centering
			\includegraphics[width=\columnwidth, trim=2cm 0cm 18cm 0.75cm, clip]{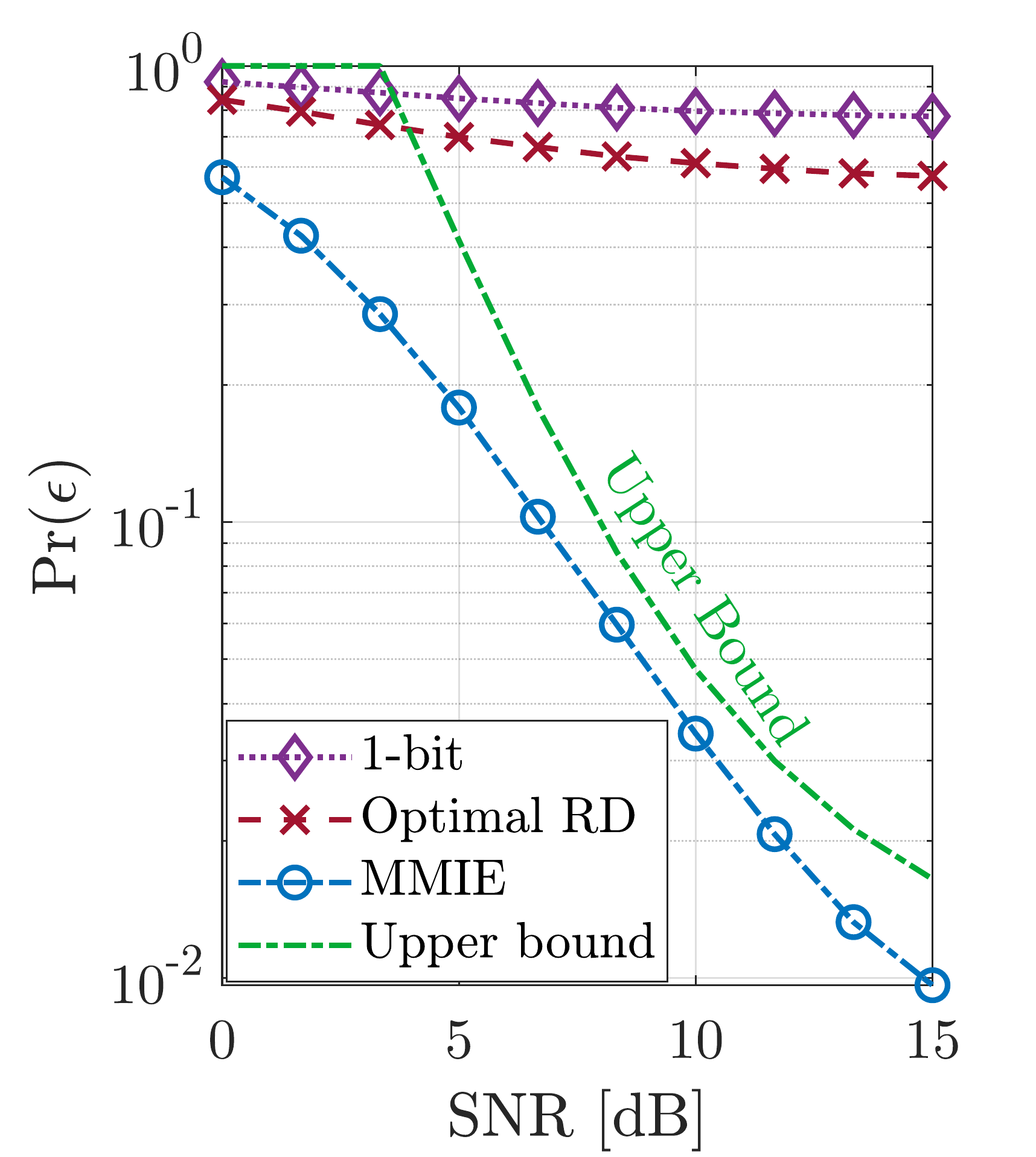}
			\caption{}
			\label{fig:subfig2}
		\end{subfigure}
		\vspace{-0.3cm}
		\caption{Error probability vs.\ (a) the message size for $\snr=20$dB; and (b) the SNR for $k=15$ bits. The black dashed line in (a) is $\widehat{c}\cdot 2^{-k\frac{|\rho|^2}{2-|\rho|^2}}$, where $\widehat{c}$ is the best least-squares-fitted constant for the MMIE's curve. The green dashed line in (b) is the minimum of \eqref{eq:upperbound} and $1$.}
		\label{fig:simulationresults}
		\vspace{-0.6cm}
	\end{figure}

	We generate the signal according to model \eqref{eq:model} with $d_{m}=150$ fixed, and compute (i) the MMIE \eqref{eq:maxestsimple}; (ii) the CCE \eqref{eq:ccedefinition} with $\rndx[n]$ replaced by $\widehat{\rndx}_{\ratedist}[n]$,\footnote{Since $\rndx[n]\overset{\text{iid}}{\sim}\jpg(0,1)$, its real and imaginary parts are statistically independent white Gaussian processes (WGP). Hence, we compute the resulting optimal distortion by allocating $\tfrac{k}{2}$ bits to the real and imaginary parts, which are independent WGP, and create their noisy versions accordingly \cite[Ch.~10.3.2]{cover1999elements}.} a RD-optimally compressed version thereof, with the squared error $\left(\rndx-\widehat{\rndx}\right)^2$ as the distortion measure, in the sense that an optimal RD compression is applied to the real and imaginary parts individually; and (iii) the CCE \eqref{eq:ccedefinition} when $\rndx[n]$ is replaced by $\widehat{\rndx}_{\onebit}[n]\triangleq\sign\left(\Re\left\{\rndx[n]\right\}\right)+\jmath\sign\left(\Im\left\{\rndx[n]\right\}\right)$. Figure~\ref{fig:subfig1}, which shows $\prob{\epsilon}$ vs.\ $k$ for $\snr=20$dB, reflects a good empirical fit to our result \eqref{eq:exactasymptoticerrorexponent}, and further demonstrates how our method outperforms any compression scheme that opts to compress a subsequence of $\rndx[n]$ in the MSE sense while ignoring the existence of $\rndy[n]$, and in particular $1$-bit per sample scalar quantization. In Fig.~\ref{fig:subfig2}, we show the error probability vs.\ the SNR, when the message size $k=15$ bits is fixed. The upper bound is still informative at moderate SNR levels.
	
	We therefore see that the proposed extremum-encoding compression technique, with our adapted time-delay estimator, can be useful and attractive for the Gaussian complex-valued signal model, which more naturally lends itself to a baseband signal model than the one considered in \cite{weiss2024joint}, and therefore to other domains (e.g., the RF domain).
	\vspace{-0.65cm}
	\section{Concluding Remarks}\label{sec:conclusion}
	\vspace{-0.2cm}
	We extended our recently proposed joint compression-TDE scheme for distributed systems with communication constraints to the important complex-valued case. We analyze its asymptotic performance, and specifically derive its exact error exponent. Our scheme is intuitive, simple to implement, and is universal in the sense that it is agnostic to the noise level, hence it does not require prior knowledge of the SNR.
	
	Current research efforts are focused on a refined, non-asymptotic analysis (for this and other regimes of operation) and additional extensions, e.g., including a Doppler effect for joint time-delay and frequency-difference of arrival estimation. Ultimately, our proposed scheme suggests that there is a considerable potential for improving communication efficiency in the context of this ubiquitous task, and beyond.
	
	\bibliographystyle{IEEEbib}
	\bibliography{refs}
	
\end{document}